\documentclass[preprint,5p,numbers,sort&compress]{elsarticle}

\usepackage{hyperref}
\usepackage{graphicx}
\usepackage{amsmath}
\usepackage{mathrsfs}
\usepackage{xspace}
\usepackage{color}

\newcommand{\eq}[1]{Eq.~\eqref{eq:#1}}

\newcommand{\mae}[3]{\langle#1\rvert#2\rvert#3\rangle}

\newcommand{\ket}[1]{\lvert#1\rangle}

\newcommand{\df}{\mathrm{d}}

\def\cE{\mathcal{E}}

\newcommand{\si}{\sigma}

\newcommand{\nn}{\nonumber}

\newcommand{\Pythia}{\textsc{Pythia}}

\allowdisplaybreaks[2]

\begin{document}

%%%%%%%%%%%%%%%%%%%%%%%%%%%%%%%%%%%%%%%%%%%%%%%%%%%%%%%%%%%%%%%%%%%%%%%%%%%%%%%%
% Title page
%%%%%%%%%%%%%%%%%%%%%%%%%%%%%%%%%%%%%%%%%%%%%%%%%%%%%%%%%%%%%%%%%%%%%%%%%%%%%%%%

\begin{frontmatter}

\title{Putting Jet Substructure  on Track(s)}

\author[Yale,ANL]{Kyle Lee}
\ead{kyle@anl.gov}

\author[Yale]{Ian Moult}
\ead{ian.moult@yale.edu}

\author[UvA,Nikhef]{Wouter J.~Waalewijn}
\ead{w.j.waalewijn@uva.nl}

\affiliation[Yale]{organization={Department of Physics, Yale University},
                  city={New Haven},
                  state={CT},
                  postcode={06511},
                  country={USA}}

\affiliation[ANL]{organization={Physics Division, Argonne National Laboratory},
                 city={Lemont},
                 state={IL},
                 postcode={60439},
                 country={USA}}

\affiliation[UvA]{organization={Institute of Physics, University of Amsterdam},
                 addressline={Science Park 904},
                 city={Amsterdam},
                 postcode={1098 XH},
                 country={The Netherlands}}

\affiliation[Nikhef]{organization={Nikhef, Theory Group},
                    addressline={Science Park 105},
                    city={Amsterdam},
                    postcode={1098 XG},
                    country={The Netherlands}}

%%%%%%%%%%%%%%%%%%%%%%%%%%%%%%%%%%%%%%%%%%%%%%%%%%%%%%%%%%%%%%%%%%%%%%%%%%%%%%%%
\begin{abstract}

One of the main advances in analysis strategies at the Large Hadron Collider (LHC) has been the ability to study the detailed structure of energy flow within high transverse momentum jets, a field referred to as jet substructure. 
Jet substructure has provided new ways to search for new physics, measure Standard Model parameters, and study the dynamics of the strong nuclear force.
To push to the next level of precision, and to make measurements of increasingly subtle correlations, requires exquisite angular resolution achieved through the use of tracking information. 
In this paper we leverage recent progress in our understanding of factorization theorems and renormalization group techniques to present the first complete calculations of jet substructure observables at the LHC on tracks. 
We compute projected energy correlators up to four points at next-to-leading collinear logarithmic accuracy, matching the state of the art for jet substructure observables, but extending to tracks.
This marks a significant step in enhancing the collider physics program, enabling precise and systematically improvable comparisons between experimental measurements and theoretical calculations, made possible by the exceptional angular resolution of tracking.

\end{abstract}
%%%%%%%%%%%%%%%%%%%%%%%%%%%%%%%%%%%%%%%%%%%%%%%%%%%%%%%%%%%%%%%%%%%%%%%%%%%%%%%%

\end{frontmatter}

%%%%%%%%%%%%%%%%%%%%%%%%
\section{Introduction}

Colliders provide the most powerful means of probing nature at the shortest accessible length scales. Details of the underlying microscopic collision are encoded in macroscopic correlations in the energy flux of emerging particles, much like how primordial fluctuations are encoded in the cosmic microwave background (CMB). The goal of collider physics is to decode these correlations to infer properties of the underlying microscopic dynamics. While this idea is straightforward in principle, it is complicated in practice due to the nontrivial mapping between the microscopic particles of the collision and the hadrons observed in detectors. This is particularly true in the complex environment of hadronic collisions.

\begin{figure}[!t]
\begin{center}
\includegraphics[width=0.49\textwidth]{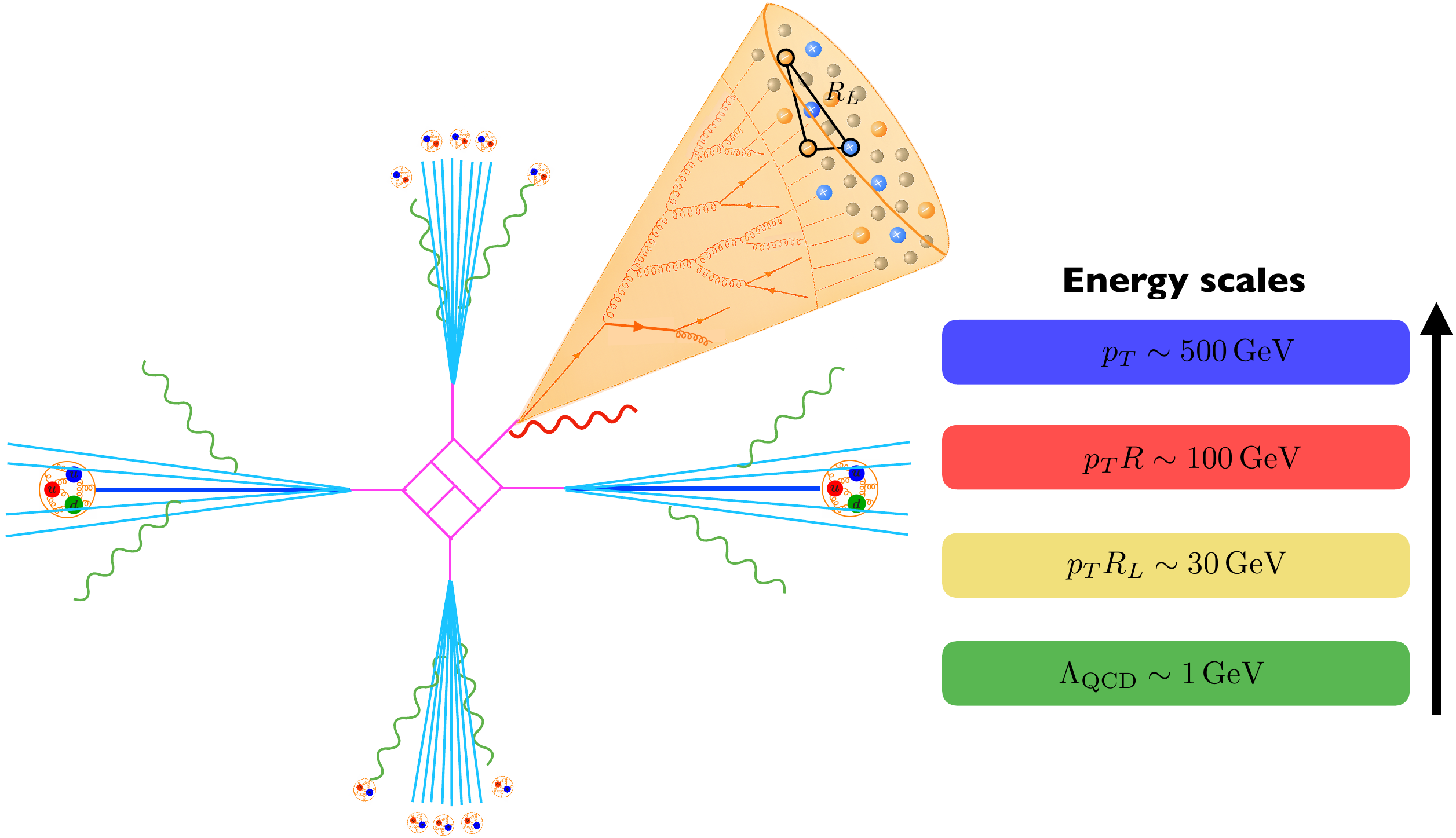}
\end{center}
\caption{Illustration of the measurement of the track-based projected three-point energy correlator on an inclusive jet sample, produced in proton-proton collisions at the LHC. The incorporation of tracking information induces a dependence on dynamics at the scale $\Lambda_{\text{QCD}}$, significantly complicating the theoretical description.
}
\label{fig:illustration}
\end{figure}

Significant progress in the theoretical understanding and experimental measurement of
these correlations has been achieved at the LHC, with the development of jet substructure \cite{Larkoski:2017jix,Kogler:2018hem}. This has enabled the measurement and calculation of correlations in the energy flux within individual jets, and has seen a wide range of applications from searches for new physics to measurements of fundamental parameters and studies of QCD dynamics. 

In the last several years there has been a program \cite{Moult:2025nhu} to systematically reformulate the study of jet substructure in terms of correlation functions of energy flow operators \cite{Sterman:1975xv,Sveshnikov:1995vi,Tkachov:1995kk,Korchemsky:1999kt}, so-called energy correlators. Building on their original application in $e^+e^-$ colliders \cite{Basham:1978zq,Basham:1979gh,Basham:1977iq,Basham:1978bw}, they were extended to jet substructure in proton-proton collisions \cite{Komiske:2022enw}, nucleus-nucleus collisions \cite{Andres:2022ovj}, and electron-proton collisions \cite{Liu:2022wop}. This program has had diverse experimental successes, including precision measurements of the strong coupling constant in proton-proton collisions \cite{CMS:2024mlf}, the observation of medium modification in proton-nucleus \cite{ALICE:2026giw,ALICE:2026zyx} and nucleus-nucleus collisions \cite{CMS:2025ydi}, and the measurement of single spin asymmetries \cite{STAR:2026epw}. For a review, and a more detailed selection of references, see \cite{Moult:2025nhu}.

The continued success of this program relies on the ability to measure correlators with increasingly fine angular resolution, in complicated collision environments, as well as to measure higher-point correlators, analogous to the study of non-gaussianities in the CMB \cite{Chen:2019bpb,Chen:2022swd,Bossi:2024qho,Barata:2025fzd}. Physics applications that will benefit from these improvements include studies of the confinement transition \cite{Komiske:2022enw,Lee:2025okn,Chang:2025kgq,Herrmann:2025fqy}, searches for nuclear modification in nuclear collisions~\cite{Andres:2022ovj,Barata:2023bhh,Andres:2023ymw,Andres:2023xwr,Yang:2023dwc,Andres:2024xvk,Andres:2024ksi,Bossi:2024qho,Barata:2025fzd}, and measurements of the top quark mass \cite{Holguin:2022epo,Holguin:2024tkz,Holguin:2023bjf}. Experimentally, such measurements will require exceptional angular resolution, which can be achieved using tracking information.  The advantages of track-based measurements for energy correlators have been illustrated in a variety of collision systems, including archival ALEPH \cite{Electron-PositronAlliance:2025fhk,Electron-PositronAlliance:2025wzh} and DELPHI \cite{Zhang:2025nlf} $e^+e^-$ data, proton-proton collisions \cite{ALICE:2024dfl,ALICE:2025igw,STAR:2025jut},
proton-nucleus collisions \cite{ALICE:2026giw,ALICE:2026zyx}, heavy ion collisions \cite{CMS:2025ydi}, and measurements of higher point correlators in proton-proton collisions \cite{Chen:2022swd,Komiske:2022enw}.

Since track-based measurements depend on quantum numbers of final state hadrons, namely their electric charge, they are sensitive to details of the confinement transition. This significantly complicates the theoretical description, as it necessitates theoretical control from the multi-TeV scale of the underlying microscopic collision, to the GeV scale of confinement, as illustrated in Fig.~\ref{fig:illustration}. This is achieved through rigorous QCD factorization theorems~\cite{Collins:1981ta,Bodwin:1984hc,Collins:1985ue,Collins:1988ig,Collins:1989gx,Collins:2011zzd,Nayak:2005rt}, advanced by effective field theory methods~\cite{Bauer:2000ew, Bauer:2000yr, Bauer:2001ct, Bauer:2001yt,Bauer:2002nz,Beneke:2002ph,Rothstein:2016bsq}, which enable a precision description of the dynamics at disparate scales, and their subsequent connection using renormalization group techniques.

\begin{figure}[!t]
\begin{center}
\includegraphics[width=0.48\textwidth]{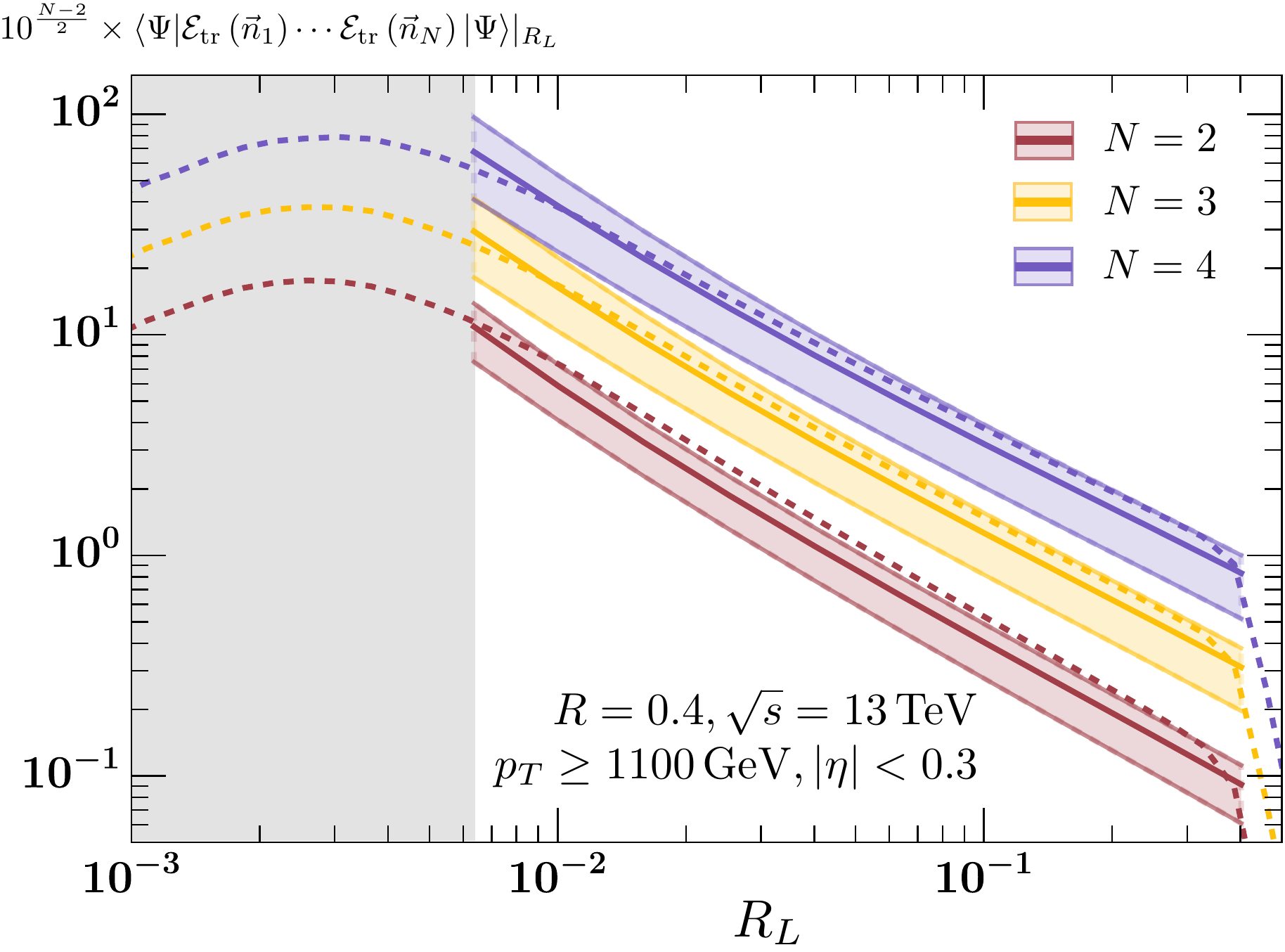}
\end{center}
\caption{Projected $N$-point energy correlators on tracks at NLL (solid), compared to \Pythia~(dashed), for LHC kinematics. The bands show the perturbative uncertainty and uncertainty from $\bar \Omega_i^{\rm tr}$, as described in the text. Our calculation does not describe the confinement transition (gray).}
\label{fig:ENC}
\end{figure}

A seminal advance toward incorporating tracking information in theoretical calculations was the introduction of track functions~\cite{Chang:2013rca,Chang:2013iba}, universal non-perturbative objects that describe the fraction of energy carried by  \emph{all} charged hadrons from a fragmenting quark or gluon. By contrast, standard fragmentation functions encode the energy fractions of \emph{individual} hadrons. Track functions are thus the appropriate non-perturbative matrix elements to describe correlations in energy flux relevant for jet substructure. Compared to hadronization models in parton showers, they are systematically improvable, and the ingredients at $\mathcal{O}(\alpha_s^2)$ have recently been obtained \cite{Chen:2022pdu,Li:2021zcf,Jaarsma:2022kdd,Chen:2022muj}, making high precision possible. These techniques have been illustrated in the context of energy correlators in $e^+e^-$ colliders, where they enabled calculations to state-of-the-art accuracy, extended to tracks \cite{Electron-PositronAlliance:2025fhk,Jaarsma:2025tck}. However, due to the complexity of jet substructure measurements in a hadron collider environment, it has been a longstanding challenge to achieve complete theoretical predictions of track-based jet substructure observables at the LHC.

In this paper, we present first-principles predictions for projected $N$-point energy correlators \cite{Chen:2020vvp} computed on tracks in an inclusive LHC jet sample, defined as
%%%
\begin{align} \label{eq:penc_def}
     \frac{\df \sigma^{[N],\rm tr}}{\df p_T\, \df \eta\, \df R_L} &= \int \df \si_{\rm jet} \sum_{i_1, \dots i_N \in \rm tr} \frac{p_{T,i_1} \cdots p_{T,i_N}}{p_T^N} 
     \nn \\ & \quad \times
     \delta\bigl(R_L - \max\{\Delta R_{i_j,i_k}\}\bigr)
\,.\end{align}
%%%
Here $\df \sigma_{\rm jet}$ denotes the inclusive jet cross section with transverse momentum $p_T$ and rapidity $\eta$. The sums on $i_1, \dots, i_N$ run over all \emph{charged} constituents of the jet, with $p_{T,i_j}$ the transverse momentum of $i_j$, and the distance between $i_j$ and $i_k$ in $(\eta,\phi)$-space denoted by $\Delta R_{i_j,i_k}$. The delta function encodes that $R_L$ equals the largest pair-wise distance. Note that this cross section does not satisfy the usual sum rule for energy correlators, since the denominator involves the transverse momentum of all particles in the jet, which avoids distortions from jets with few charged particles. We use $\left.\mae{\Psi}{\cE_{\rm tr}(\vec n_1)\cdots\cE_{\rm tr}(\vec n_N)}{\Psi}\right|_{R_L}$ as shorthand for the normalized track-based projected correlator in Eq.~\eqref{eq:penc_def}, with $\ket{\Psi}=\ket{\Psi(p_T,\eta)}$ denoting the identified jet state.

Our results are enabled by several recent theoretical advances. 
First, there has been significant progress in understanding the non-linear renormalization group equations governing the scale evolution of track functions, including the calculation of higher-order corrections~\cite{Li:2021zcf,Chen:2022pdu,Chen:2022muj,Jaarsma:2022kdd}, which is essential for precision predictions.
Second, for the class of observables known as energy correlators, the incorporation of tracks has been shown to admit a simple and systematic treatment~\cite{Chen:2020vvp,Jaarsma:2023ell,Lee:2023npz,Lee:2023tkr,Jaarsma:2025tck}.
Third, nonperturbative corrections to energy correlators, which are known to be sizable from Monte Carlo parton showers, are now under field-theoretic control for projected energy correlators~\cite{Schindler:2023cww,Chen:2024nyc,Lee:2024esz,Budhraja:2026pyi}.
Fourth, recent developments in the description of inclusive jet production have extended its accuracy beyond leading logarithms in the jet radius, enabling reliable predictions for jet substructure observables defined on inclusive jet samples~\cite{vanBeekveld:2024jnx,Lee:2024icn,Lee:2024tzc,Generet:2025vth}.

%%%%%%%%%%%%%%%%%%%%%%%%
\section{Factorizing Track Energy Correlators at the LHC}
\label{sec:factorization}

The framework we use builds on the factorization theorem for (projected) energy correlators in the small angle limit of $e^+e^-$ collisions~\cite{Dixon:2019uzg}, extended to inclusive jets at the LHC~\cite{Lee:2022uwt,Lee:2024icn}. For the description of the inclusive jet sample, higher logarithmic accuracy has recently been achieved in~\cite{Lee:2024tzc,Generet:2025vth}. 
To extend this to track-based predictions, we use results for energy correlators in $e^+e^-$ collisions developed in~\cite{Chen:2020vvp,Jaarsma:2023ell}.

\begin{figure}[!t]
\begin{center}
\includegraphics[width=0.48\textwidth]{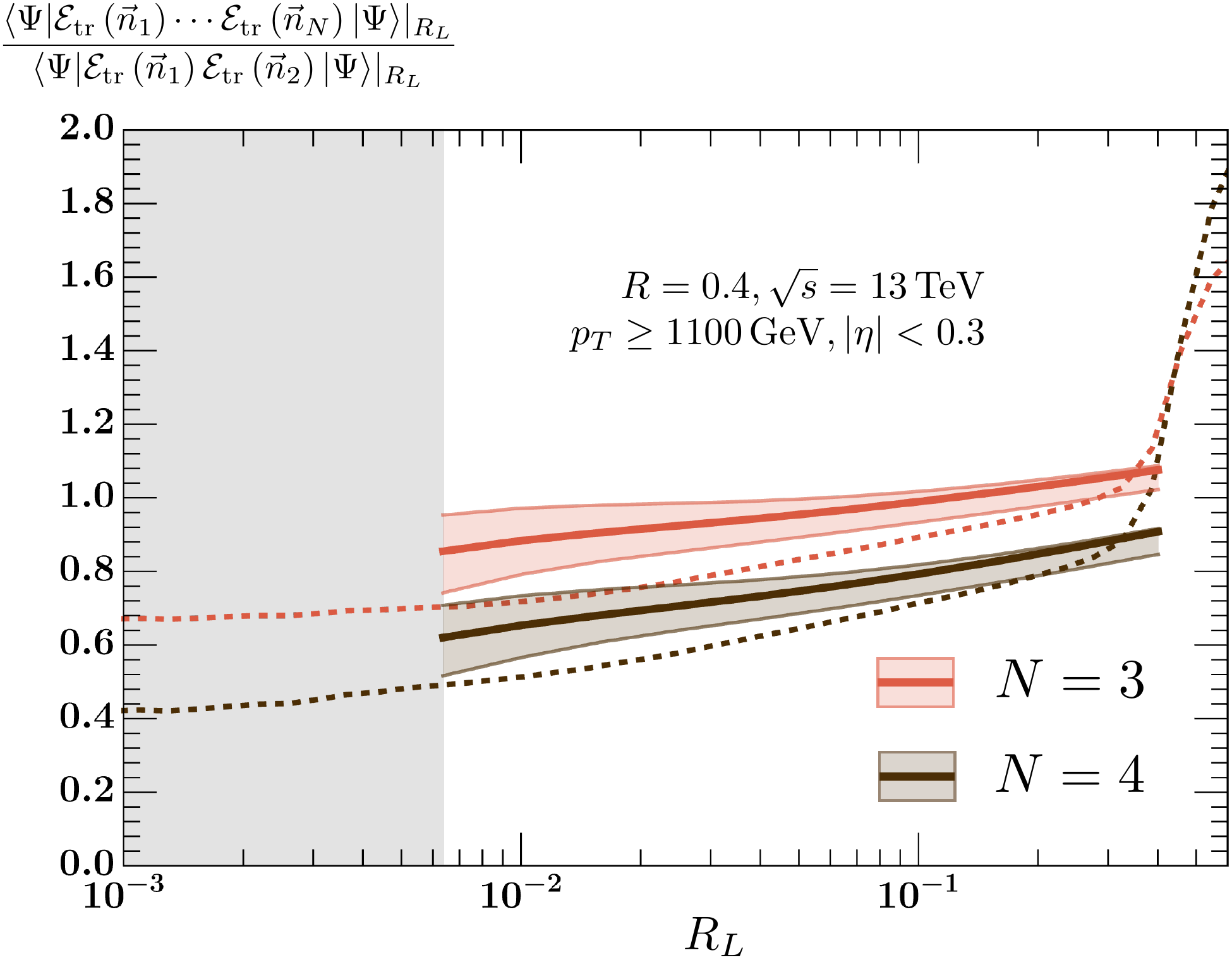}
\end{center}
\caption{Projected $N$-point energy correlator divided by two-point correlator, both on tracks, at NLL (solid) and from \Pythia~(dashed). Uncertainties from scale variations and $\bar \Omega_1^{\rm tr}$ are treated as correlated in the ratio. Our calculations do not extend into the confinement transition (gray).}
\label{fig:ENCtoE2C}
\end{figure}

We now describe the framework we use to calculate the track-based energy correlator in LHC jets. The starting point is the factorization~\cite{Lee:2024icn,Generet:2025vth} for the cross section cumulative in $R_L$
%%%
 \begin{align} \label{eq:fac}
  &\frac{\df \Sigma^{[N],\rm tr} (R_L)}{\df p_T\, \df \eta} \equiv \int^{R_L} \! \df R_L'\,\frac{\df \sigma^{[N],\rm tr}}{\df p_T\, \df \eta\, \df R_L'} 
   \\ & \quad
  = \sum_i \int \df z\, \mathcal{H}_i(p_T/z,\eta,\mu) 
  \nn \\ & \qquad \times
  \int\! \df x\, x^N \mathcal{J}_{ij}(z,p_T R,x,\mu)\,   J_j^{[N],\rm tr}(x p_T R_L, \mu)
\,.\nn \end{align}
%%%
We now describe its ingredients, starting with the production ($\mathcal{H}_i$) of a hard parton $i$ with transverse momentum $p_T/z$ and rapidity $\eta$ (including parton distributions for the incoming protons). This in turn results in a jet ($\mathcal{J}_{ij}$)  initiated by parton $j$  with transverse momentum $p_T$ and radius parameter $R$, on which the track-based projected $N$-point energy correlator ($J_j^{[N], \rm tr}$) with the largest distance $R_L$ is measured. Fig.~\ref{fig:illustration} illustrates each of these dynamics described by the factorization ingredients.

The factorization in \eq{fac} separates physics at different scales 
\begin{equation}
\mu_{\mathcal{H}} \sim p_T 
\quad \gg \quad
\mu_{\mathcal{J}} \sim p_T R 
\quad \gg \quad
 \mu_{J} \sim p_T R_L 
.\end{equation} 
Using renormalization group evolution to evolve these ingredients from their natural scales to a common scale, resums the logarithms of $\mu_{\mathcal{J}}/\mu_{\mathcal{H}} \sim R$ and $\mu_{J}/\mu_{\mathcal{J}} \sim R_L/R$.
We will vary each of these scales by a factor of 2 to assess the perturbative uncertainty, taking their envelope.
The explicit expressions we need are given in~\cite{Aversa:1988mm, Aversa:1988fv, Aversa:1988vb, Aversa:1989xw, Aversa:1990uv, Jager:2004jh} for $\mathcal{H}_i$,~\cite{Kang:2016ehg} for $\mathcal{J}_{ij}$ and~\cite{Jaarsma:2025tck} for $J_j^{[N],\rm tr}$. We use the parton distributions from the CT14NLO PDF set~\cite{Dulat:2015mca}.

\begin{figure}[!t]
\begin{center}
\includegraphics[width=0.48\textwidth]{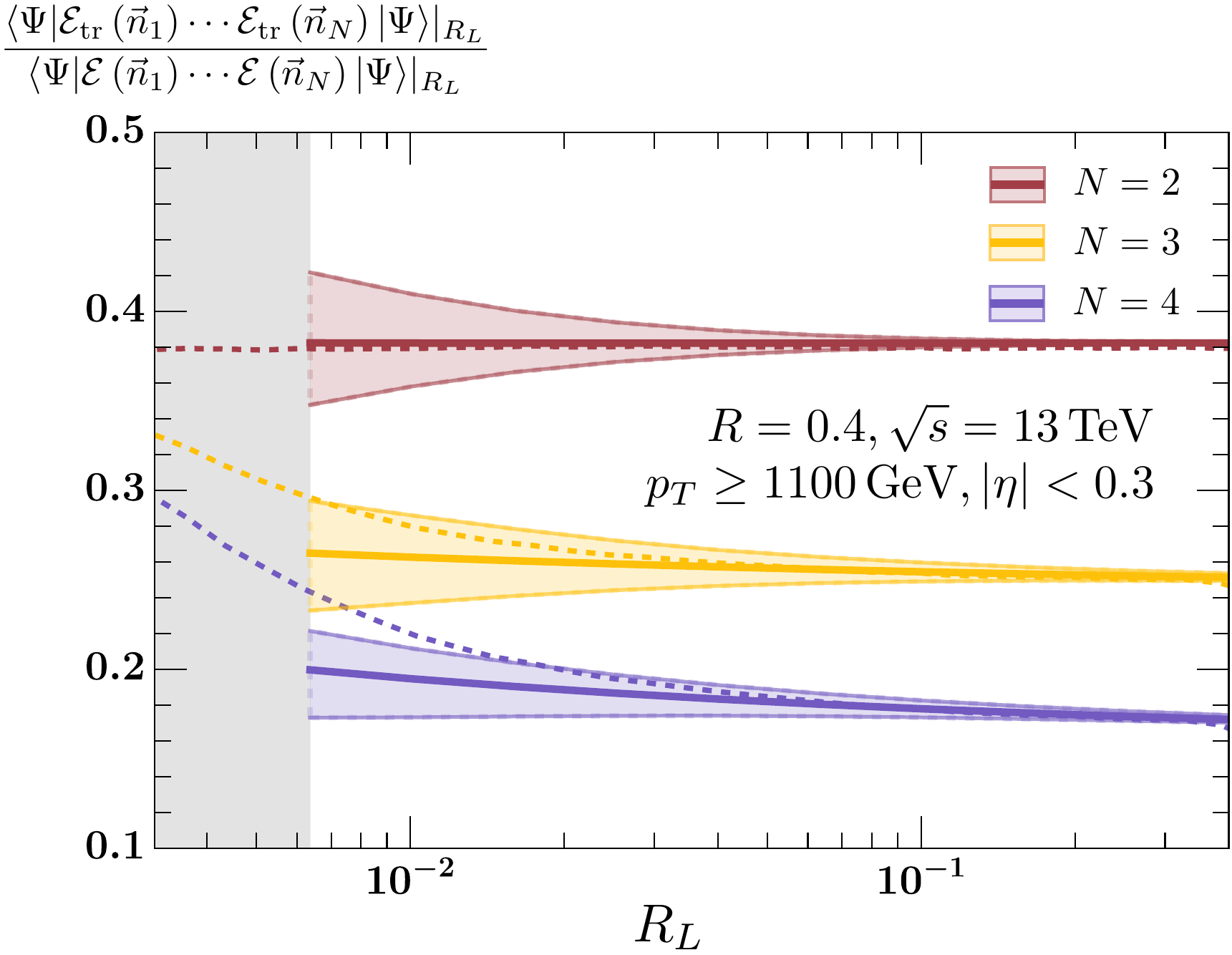}
\end{center}
\caption{Projected $N$-point energy correlator on tracks divided by the corresponding correlator on all particles, at NLL (solid) and from \Pythia~(dashed). Uncertainties from scale variations and $\bar \Omega_1$ are treated as correlated in the ratio.}
\label{fig:chtoall}
\end{figure}
The jet function for projected energy correlators on tracks, $J_j^{[N], \rm tr}$, encodes the projected energy correlator measurement, involving perturbative physics at the scale $p_T R_L$ and the conversion to tracks at the scale $\Lambda_{\rm QCD}$. The physics at these disparate scales is incorporated by matching onto moments of track functions. At lowest order,  $\hat J_j^{[N],\rm tr}(x p_T R_L, \mu) = T_j(N,\mu)$. At higher orders, it involves logarithms of $x p_T R_L/\mu$ and up to $N$ track function moments. The track function moments we take as input are from~\cite{Chang:2013rca}. 

The leading nonperturbative corrections to this matching are given by
%%%
\begin{align}
  J_j^{[N],\rm tr}(x p_T R_L,\mu) &= \hat J_j^{[N],\rm tr}(x p_T R_L,\mu) 
   \\ & \quad 
  - \frac{N \bar \Omega_{1j}^{\rm tr}}{x p_T R_L} \mathcal{J}_j^{[N-1],\rm tr} (x p_T R_L,\mu) 
\,,\nn\end{align}
%%%
where $\hat J_j^{[N],\rm tr}$ is the perturbative contribution, $\mathcal{J}_j^{[N-1],\rm tr}$ is the perturbative matching coefficient for the leading nonperturbative power correction, and $\bar \Omega_{1j}^{\rm tr}$ are nonperturbative matrix elements. At tree-level, $\mathcal{J}_j^{[N-1],\rm tr} = T_j(N-1,\mu)$ and the logarithmic terms at higher orders are fixed by RG consistency. 
We take $\bar \Omega_{1,q} = 0.305\,\rm{GeV}$ from~\cite{Schindler:2023cww} and approximate $\bar \Omega_{1,g} = (C_A/C_F) \bar \Omega_{1,q}$ varying this by $\pm 30$\% as part of our uncertainty estimate. 
To account for the conversion to tracks, we use isospin symmetry to approximate $\bar \Omega_{1j}^{\rm tr} \approx 2/3 \bar \Omega_{1j}$, varying this by 10\% to assess the corresponding uncertainty. We simply combine these nonperturbative variations and perturbative scale variations as a total envelope. There are also subleading $\mathcal{O}(\Lambda_{\rm QCD}/p_TR_L)$ nonperturbative corrections to the matching of the jet function $J_j$ onto track functions, that do not have the  $1/R_L^2$ enhancement in the collinear limit. 

Finally, the factorization in \eq{fac} receives power corrections from the expansion in the small-$R$ and small-$R_L$ limits. The leading corrections are $\mathcal{O}(R^2)$ and $\mathcal{O}[(R_L/R)^2]$. They are not included in our results, but could be incorporated through matching. The $\mathcal{O}[(R_L/R)^2]$ power corrections can be added by matching onto a calculation in which $R \sim R_L$, and the $\mathcal{J}_{ij}$ and $J_j^{[N],{\rm tr}}$ are no longer factorized but described
by one function. Likewise, the $\mathcal{O}(R^2)$ corrections can be incorporated by matching onto a fixed-order calculation in which the hard and jet scales are not factorized. These power corrections were recently studied in \cite{Generet:2025vth}, and found to be numerically small for the kinematics considered in this paper.

%%%%%%%%%%%%%%%%%%%%%%%%
%%%%%%%%%%%%%%%%%%%%%%%%
\begin{figure}[!t]
\begin{center}
\includegraphics[width=0.48\textwidth]{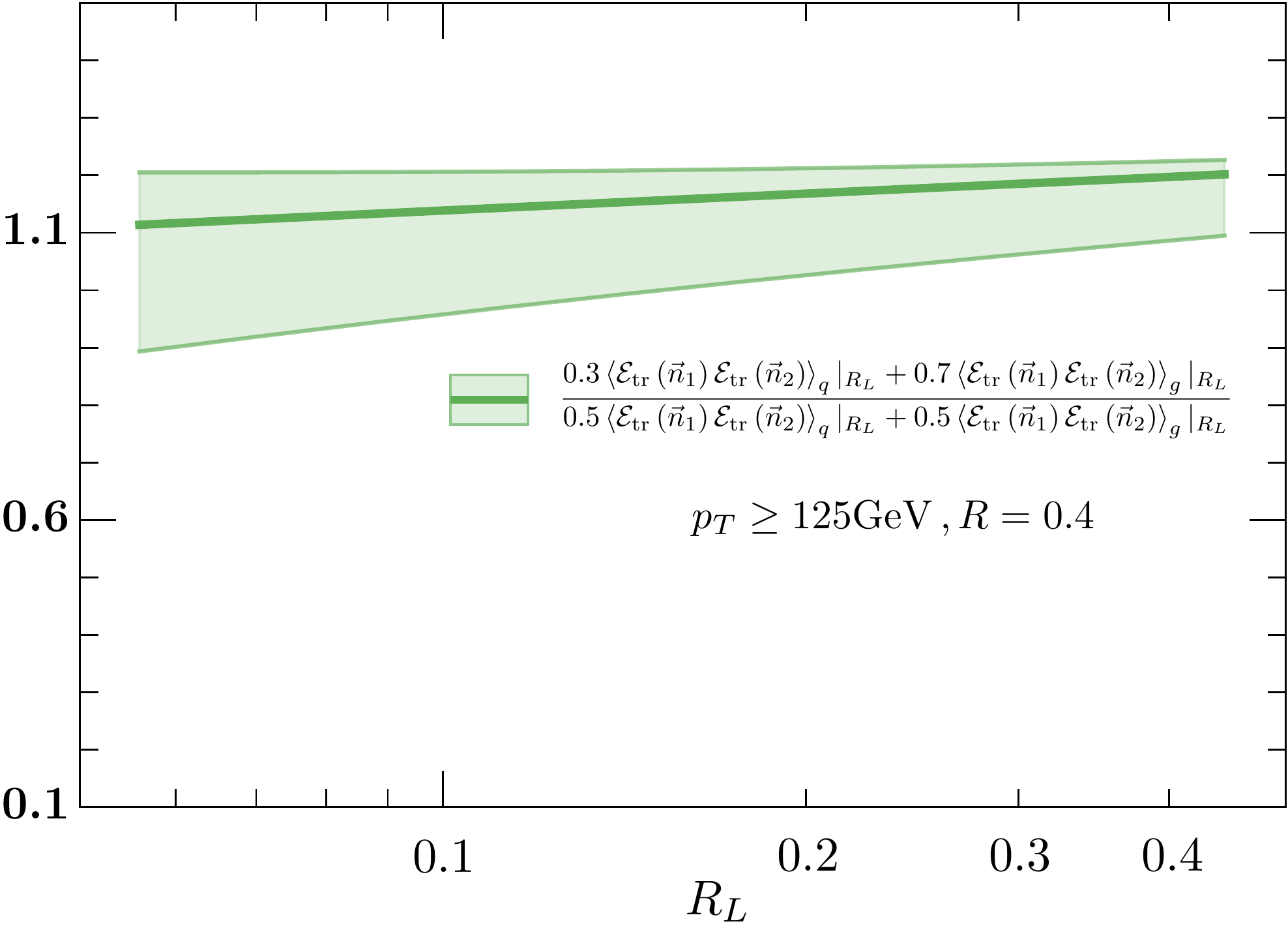}
\end{center}
\caption{The two-point energy correlator on tracks for different mixtures of quark and gluon jets (30/70 vs.~50/50) at NLL. Uncertainties from scale variations and $\bar \Omega_1^{\rm tr}$ are treated as correlated in the ratio.}
\label{fig:modifiedqg}
\end{figure}

\section{Results for Projected Track Energy Correlators}

We now describe our results, starting with predictions for the track-based projected $N$-point energy correlator in Fig.~\ref{fig:ENC}. Here we show our predictions at NLL accuracy, with the band accounting for perturbative uncertainties of each of the ingredients (through their scale variations) and the uncertainty on $\Omega_i^{\rm tr}$, as described in Sec.~\ref{sec:factorization}. Although the nonperturbative is power suppressed, it still has a substantial effect because its classical scaling is $1/R_L^{2}$ compared to $1/R_L$ for the perturbative contribution. The grayed out region of the plot corresponds to the confinement transition, which is not described by our calculation. To ensure a sufficiently large perturbative range in $R_L$, we choose LHC kinematics with a large jet transverse momentum. We compare our results to \Pythia~8.3~\cite{Bierlich:2022pfr}, finding reasonable though not perfect agreement. We also note that the largest difference is for $N=2$ and concerns the normalization rather than the shape of the distribution. Going to NNLL accuracy of the jet production~\cite{Generet:2025vth} is critical for getting the right normalization relative to data.

Next, we take the ratio of these track-based projected correlators to the two-point correlator in Fig.~\ref{fig:ENCtoE2C}. Here we vary the perturbative scales and the $\Omega_i^{\rm tr}$ in a correlated manner between numerator and denominator. This is somewhat justified by their shared structure: perturbatively, both inherit their anomalous scaling from twist-two operators, differing only in spin and accessed through different moments of the same QCD splitting functions; nonperturbatively, the same $\Omega_i^{\rm tr}$ enters both. As a result, the band narrows substantially and the \Pythia~result lies outside it, though this discrepancy is again largely a normalization effect. A quantitative treatment of uncertainties for ratio observables that properly incorporates correlations between the numerator and denominator will be important for the application of our results to precision measurements. There has been some recent progress in this direction with the development of the theory nuisance parameter approach~\cite{Tackmann:2024kci}. However, its application to ratios of projected energy correlators is beyond the scope of this paper, and we leave it to future work.

In Fig.~\ref{fig:chtoall}, we take the ratio of the projected $N$-point correlator on tracks to the corresponding correlator on all hadrons. Again, we treat the numerator and denominator as correlated, although the uncertainty due to the relation between $\Omega_i^{\rm tr}$ and $\Omega_i$ of course only affects the numerator. The two-point correlator is almost independent of $R_L$, while the higher-point correlators have a nontrivial shape, in line with~\cite{Jaarsma:2023ell}. The dominant uncertainty in this plot is due to nonperturbative effects, which explains the growth in the uncertainty band at small $R_L$. 
Indeed, varying the nonperturbative parameters in our calculation can improve the agreement with \Pythia, motivating a dedicated fit to determine these parameters more precisely.

Finally, we explore the dependence on the quark/gluon composition of jets in Fig.~\ref{fig:modifiedqg}. This is of particular interest for measurements of energy correlators in proton-nucleus (pA) collisions~\cite{ALICE:2026giw,ALICE:2026zyx}, where nuclear effects modify the quark/gluon composition of jets, enhancing the gluon fraction. Disentangling such a composition change from genuine modifications of the jet substructure itself due to nuclear modification to the partonic splitting is an important question, and the ratio we present can loosely be viewed as a proxy for $R_{pA}$ that accounts only for quark/gluon-fraction modifications. Concretely, we take the ratio of the two-point correlator for a 30/70 and a 50/50 composition, at $\mu_\mathcal{J}\sim p_T R$ scale~\cite{Cal:2019hjc,Cal:2020flh} and again treat the uncertainties as correlated. Note that we also take $p_T\geq 125\,\rm{GeV}$, which are more realistic $p_T$ values for collisions involving nuclei. The result is consistent with 1 over most of the range, with a slight enhancement at large $R_L$, but the band is large, showing that the quark/gluon fraction alone can drive a fair amount of variation. A reliable computation of this large-$R_L$ enhancement will be important, since such features must be disentangled from the genuine nuclear modifications of the partonic splitting discussed above, which also modify the correlator at large $R_L$. A true first-principles computation for pA, integrating the appropriate nuclear parton distribution functions, would be an important next step. At these lower $p_T$, going to even higher order in perturbative accuracy would also be important to reduce the uncertainty.

\section{Conclusions}
Jet substructure at the LHC is an intrinsically multi-scale problem, spanning the hard scale of jet production, the collinear dynamics within the jet, and the nonperturbative scales of hadronization and confinement. The quest to achieve precision jet substructure physics in such a complicated LHC environment has been long. It has required new ways of thinking about the observables used for jet substructure measurements, the development of new approaches for proving factorization theorems, and improvements in our understanding of renormalization group equations for Lorentzian field theory.

In this paper we have combined all these ingredients to provide complete, first-principles predictions for jet substructure observables on tracks at the LHC, representing a significant step in the program of achieving precision jet substructure at hadron colliders. Our calculations are crucial for comparison with numerous hadron collider measurements of energy correlators, which have been performed on tracks \cite{ALICE:2024dfl,CMS:2024mlf,CMS:2025ydi,STAR:2025jut}. 

While our results illustrate that it is possible to combine all the recent progress in the calculation of energy correlators, this is merely a point of departure. All of the ingredients required to advance these predictions by one order in perturbative accuracy are now available~\cite{Chen:2022muj,Lee:2024icn,Generet:2025vth,Lee:2026zyl}, and we expect this to be accomplished in the near future. This would  bring jet substructure at hadron colliders to the state-of-the-art accuracy currently achieved in $e^+e^-$ collisions~\cite{Dixon:2019uzg,Jaarsma:2025tck,Lee:2026zyl}. Such control over the QCD dynamics enables a broad program of applications: precision measurements of the strong coupling constant $\alpha_s$, studies of confinement dynamics, precision measurements of jet substructure in $pA$ collisions to probe nuclear modifications, and searches for new physics against a precisely understood QCD background~\cite{Ricci:2026fkj}. Putting jet substructure on tracks thus charts a path toward precision physics in the complicated LHC environment.

%%%%%%%%%%%%%%%%%%%%%%%%%%%%%%%%%%%%%%%%%%%%%%%%%%%%%%%%%%%%%%%%%%%%%%%%%%%%%%%%
\section*{Acknowledgments} We thank Yibei Li for collaboration during the early stages of this project. We thank Hannah Bossi, Rithya Kunnawalkam Elayavalli, Jack Holguin, YenJie Lee, and Xiaoyuan Zhang for useful discussions.
This research was supported in part by grant no.~NSF PHY-2309135 to the Kavli Institute for Theoretical Physics (KITP).
K.L.~is supported by the U.S.~Department of Energy under contract DE-AC02-06CH11357. I.M.~is supported by the DOE Early Career Award
DE-SC0025581, and the Sloan Foundation. 

%%%%%%%%%%%%%%%%%%%%%%%%%%%%%%%%%%%%%%%%%%%%%%%%%%%%%%%%%%%%%%%%%%%%%%%%%%%%%%%%
\bibliographystyle{elsarticle-num}
\bibliography{track_eec_pp.bib}{}
%%%%%%%%%%%%%%%%%%%%%%%%%%%%%%%%%%%%%%%%%%%%%%%%%%%%%%%%%%%%%%%%%%%%%%%%%%%%

\end{document}